\documentclass[%
reprint,
 amsmath,amssymb,
 aps,
]{revtex4-1}

\usepackage{physics}
\usepackage{lineno}

\usepackage[dvipdfmx]{graphicx}
\usepackage{bm}

\begin{document}


\title{Enhancing the stimulated emission of polarization-entangled photons \\ using passive optical components}

\author{Ryo Nozaki$^{1}$}

\author{Yoshiro Sato$^{1}$}

\author{Yoshitaka Shimada$^{1}$}

\author{Taku Suzuki$^{1}$}

\author{Kei Yasuno$^{1}$}

\author{Yuta Ikai$^{1}$}

\author{Wataru Ueda$^{1}$}

\author{Kaito Shimizu$^{1}$}

\author{Emi Yukawa$^{1}$}

\author{Kaoru. Sanaka$^{1}$}

  \email{E-mail address: sanaka@rs.tus.ac.jp}

\affiliation {$^{1}$ Department of Physics, Tokyo University of Science, Shinjuku-ku, Tokyo 162-8601, Japan}

\date{\today}

\begin{abstract}

Bright sources of polarization-entangled photon pairs are essential components for quantum information technologies. In many methods proposed so far, it is necessary to introduce a resonator that combines active optical components such as an electric optical modulator to enhance the stimulated emission of polarization-entangled photons. In these methods, however; it is technically difficult to perform the time series operation to output the stimulated entangled photons in the resonator by synchronizing laser pulses. In this paper, we propose a scheme to scale up the stimulated emission of polarization-entangled photon pairs using a resonator with only passive optical components. We show the theoretical aspects of the scheme and also perform a proof-of-principle experimental demonstration of the scheme in a double-pass configuration.  

\end{abstract}

\pacs{Valid PACS appear here}
\maketitle


\section{Introduction}

Spontaneous parametric down-conversion  (SPDC) provides to make a readily available photonic source of polarization-entangled photon pairs to demonstrate quantum information technologies such as quantum teleportation \cite{bouwmeester97}, quantum key distribution \cite{yin17}, quantum metrology \cite{ono13}, quantum holography \cite{defienne20}, and the recently developed boson sampling machine \cite{zhong20}. Two widely used approaches to producing polarization-entangled photon pairs are a thin $\beta$-barium borate (BBO) bulk crystal with type-II phase matching \cite{kwiat95} and a sandwiched BBO crystal with type-I phase matching \cite{kwiat99}. Owing to the development of quasi-phase matching techniques, the use of periodically poled KTiOPO$_{4}$ (ppKTP) or periodically poled  LiNbO$_{3}$(ppLN) has also become a standard technique \cite{armstrong96}. Several types of entanglement sources have been developed by combining such nonlinear crystals and interferometric configurations, such as Mach-Zehnder interferometers \cite{fiorentino04}, Franson interferometers \cite{sanaka02}, and Sagnac interferometers \cite{shi04,kim06,steinlechner14}.

Among these different interferometers, the Sagnac interferometer has major advantages because its symmetric geometry of the setup allows a very phase-stable condition resulting in the generation of high-quality polarization-entangled photons. In particular, the scheme with orthogonally polarized photon pairs obtained by type-II SPDC makes it possible to separate degenerate polarization-entangled photon pairs into different optical modes without postselective detection\cite{kim06}. The scheme with ordinary type-0 or type-I SPDC requires a nonpolarizing beam splitter to separate degenerate photon pairs with a 50$\%$ probability of success\cite{shi04}, wavelength filtering to separate nondegenerate pairs of photons\cite{steinlechner14}, or spatial mode filtering to separate non-collinear down-converted photons\cite{jabir}. In contrast, a double-pass configuration with type-0 or type-I SPDC has the major advantages of a simple setup and a high emission ratio of photon pairs\cite{steinlechner13}. Bidirectional pumping to a single ppKTP crystal generates polarization-entangled photons from two sets of parallel polarized photon pairs on the collinear optical mode. A nonpolarizing beam splitter or color filters are necessary to separate polarization-entangled photon pairs conditionally. The multiple reverse process of Hong-Ou-Mandel interference enables the properties of both type-0 and type-II SPDC to be satisfied simultaneously, and makes it possible to use the largest second-order nonlinear coefficient of a nonlinear crystal to generate polarization-entangled photons and also to separate degenerate photon pairs into different optical modes with a 100$\%$ probability of success in principle\cite{terashima18, terashima19}. In this scheme, the maximum emission rate of the photon pair is limited by the coefficient number of the crystal to produce the type-0 phase matching condition. 

To further improve the emission rate, we require the stimulation of polarization entangled photons. In a former experiment, the pump laser used to produce the polarization-entangled photons passed through a nonlinear crystal twice with type-II phase matching \cite{lamas01}. This condition realized the superposition state between the first and second generated entangled photon pairs and the emission rate once from the beginning nearly four times larger than that when the pump laser passed through the crystal only once. \\

It is necessary to introduce a resonator such as a laser to further improve the emission rate by passing the pump laser more than three times through the crystals. A proposed scheme is to construct the resonator by combining active optical components such as an electric optical modulator and generated down-converted photons under non-collinear optical mode using a pulse laser \cite{simon2003}. In this scheme, it is necessary to consider the time series operation to output the stimulated entangled photons in the resonator by synchronizing the operation with the rate of pump laser pulses. Although the emission rate of stimulated photon pairs is restricted by the mirror reflectivity of the resonator without such active optical components, it is technically hard to realize enough coupling efficiency and the time series operation simultaneously in the actual setup. \\

In this paper, we propose a stimulation scheme to produce polarization-entangled photon pairs using a resonator with only passive optical components and a continuous-wave pump laser. Our scheme makes it possible to output the stimulated polarization-entangled photons from a resonator without considering the time series operation using active optical components. In addition, our scheme enables the construction of a much simpler system for generating polarization-entangled photons with a collinear spatial optical mode only. Our methods have also advantage for the easiness to scale up the stimulation scheme due to the experimental configuration using a Sagnac-like interferometer and a resonator. First, we discuss the theoretical background to reveal the advantages of our stimulation scheme using only passive optical components. Second, we show a proof-of-principle experimental demonstration of our scheme in a double-pass configuration.

\section{Theory}

\begin{figure}[h]%
\centering
\includegraphics[width=0.5 \textwidth]{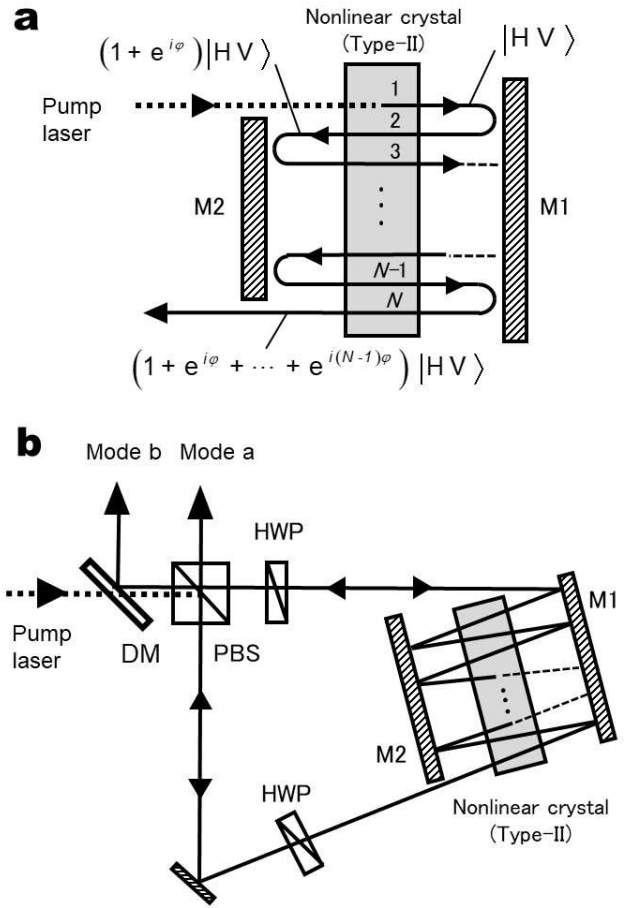}
\caption{(a) Schematic of stimulated down-converted photon pairs. orthogonally polarized photon pairs are generated in the parametric down-conversion process and are repeatedly transmitted through a nonlinear crystal between the mirrors. The phase of output photon pairs is determined by the number of times the pump beam transmits through the crystal. (b) Schematic of our double-pass configuration using a Sagnac interferometer to generate the polarization-entangled photons with stimulated down-converted photon pairs.}\label{fig1}
\end{figure}

Fig. \ref{fig1}a shows the schematic of our method of enhancing the stimulated emission of parametric down-converted photons with a resonator. A pair of photons (signal and idler photons) is produced from a single pump photon by the parametric down-conversion process, which satisfies ${\omega}_{\mathrm{p}} = {\omega}_{\mathrm{s}} + {\omega}_{\mathrm{i}}$, where ${\omega}_{\mu}$ ($\mu = {\mathrm{s}}$, ${\mathrm{i}}$, or ${\mathrm{p}}$) indicates the angular frequency of the signal, idler, or pump photon, respectively.

The polarization states of the down-converted photons is orthogonally polarized under the type-II phase matching condition as horizontal (H) and vertical (V) polarizations. When we place a nonlinear crystal inside the resonator, the down-conversion process occurs several times by reflecting back the pump light with mirrors (M1, M2). The amplitude of output photon pairs from the resonator depends on the phase $\varphi$ and the number of times the pump laser pass through the crystal ($N$) under the degenerated condition. 
Here $\varphi$ is the phase obtained by the down-converted photon pairs when they move from one mirror to another. As shown in Fig. 1(b), the crystal and the resonator are set at a finite angle with respect to the pump laser so that the crystal can be pumped by both clockwise and anti-clockwise lights of a Sangac interferometer. It is possible to pump the crystal from both directions by combining the crystal inside of the resonator with a finite angle and a Sagnac interferometer as shown in Fig.  \ref{fig1}b. In the frame of reference corotating with the pump field, the Hamiltonian of the output photon pairs generated for the clockwise (CW) direction is as follows:

\begin{equation}
\label{hcw}
\hat{H}_{\rm CW }= \kappa \sum_{m=0}^{N-1} e^{i m \varphi  } \hat{a}_H^{\dag} \hat{b}_V^{\dag} + h.c., 
\end{equation}
where the pump field is approximated as a classical field and $\kappa$ represents the coupling constant. 
The operators ${\hat{a}}_P$ and ${\hat{b}}_P$ ($P = H, V$) represent the annihilation operators of the $P$-polarized photons in the modes $A$ and $B$ in Fig.~\ref{fig1} (b), respectively, 
and the relation to the vacuum state $ |  0 \rangle $ can be described as $\hat{a}_H^{\dag} |  0 \rangle = | H_{a} \rangle,  \ \hat{b}_V^{\dag} |  0 \rangle = | V_{b} \rangle$. 
These operators satisfy the commutation relations given by  
$[{\hat{c}}_P, {\hat{c}}^{\prime}_{P^{\prime}}] = 0$ and 
$[{\hat{c}}_P, {\hat{c}}^{\prime \dagger}_{P^{\prime}}] = {\delta}_{cc^{\prime}} {\delta}_{PP^{\prime}}$ with 
$c, c^{\prime} = a, b$ and $P, P^{\prime} = H, V$. 
Similarly, the Hamiltonian of the output photon pairs generated for the counterclockwise (CCW) direction is as follows: 

\begin{equation}
\label{hccw}
\hat{H}_{\rm CCW }= \kappa \sum_{m=0}^{N-1} e^{i m \varphi } \hat{a}_V^{\dag} \hat{b}_H^{\dag} + h.c. 
\end{equation}

The Hamiltonian of photon pairs output from the Sagnac interferometer is given by the linear summation of these Hamiltonians with a relative phase relation depending on the polarization state of the pump light. Here, we choose the polarization state $-45^{\circ}$. The Hamiltonian of photon pairs output from the interferometer becomes $\hat{H}_{\rm out }=\hat{H}_{\rm CW }-\hat{H}_{\rm CCW }$. The unitary transformation by the Hamiltonian is given by $\hat{U}(t) = {\rm exp} (-i \hat{H}_{\rm out } t/ \hbar )$. Therefore, the output state is given by applying the unitary operation to the vacuum state $ |  0 \rangle$ and can be described as: 

\begin{equation}
\label{pout}
|\Psi_{\rm out} \rangle \equiv \hat{U}(t)  |  0 \rangle = {\rm exp} [ -i \sum_{m=0}^{N-1} e^{i m \varphi } (\hat{a}_H^{\dag} \hat{b}_V^{\dag}-\hat{a}_V^{\dag} \hat{b}_H^{\dag} ) \tau] \ | 0 \rangle.
\end{equation}

Here, $\tau (\equiv \kappa t/\hbar)$ is the interaction parameter. The probability of producing polarization-entangled photon pairs with the Bell state $ | {\Phi}^-_1 \rangle = ( | H_a V_b \rangle -   | V_a H_b \rangle )/ \ \sqrt{2} $ becomes 

\begin{equation}
\label{pn}
P_N \equiv P_{1,N} = |\langle {\Phi}^-_1 | \Psi_{\rm out} \rangle|^2 \approx \ 2 \tau^2 \left(\frac{\sin {N \varphi /2}}{\sin {\varphi /2}} \right)^2 , 
\end{equation} 
where $P_{M,N}$ ($M \in \mathbb{N}$) represents the probability of generating $2M$-photon maximally entangled state, 
$\tau \ll O(1)$, and $N \sim O(1)$. 
A detailed derivation of $P_{M,N}$ in Eq.~(\ref{pn}) is provided in Appendix A. 

The maximum of the probability $P_N$ becomes $2 \tau^2 N^2$ in the limit of $\varphi \rightarrow 0$. Eq. (\ref{pn}) shows that the probability of generating the polarization-entangled photon pair can be increased in proportion to the square of the number of pump light transmitted to the crystal in the resonator. In general, the probability of generating entangled photons with an even number of entangled photons $2M (M=1,2,3 \cdots)$ can be given by $(M+1) \tau^2 N^{2M}$ (Refer to Appendix A).

\section{Experiment}

\begin{figure}[h]%
\centering
\includegraphics[width=0.5 \textwidth]{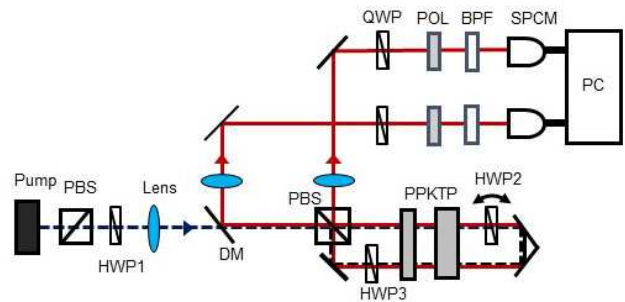}
\caption{Setup for our  proof-of-principle experimental demonstration of stimulated down-converted photons in a double-pass configuration.  The polarization of the pump laser is adjusted by a polarizing beam splitter (PBS) and a half-wave plate (HWP1). Orthogonally polarized photon pairs are generated by a ppKTP crystal in a Sagnac interferometer composed of a PBS and half-wave plates (HWP2, HWP3).  A dichroic mirror (DM) separates the pump laser and down-converted photons. The output photons are reflected by the PBS and a DM and counted by single-photon counting module (SPCMs) after passing through quarter-wave plates (QWPs), polarizers (POLs), and band-pass filters (BPFs).}\label{fig2}
\end{figure}

Fig.  \ref{fig2} shows our setup for a proof-of-principle experimental demonstration of  stimulated down-converted photons.  The polarization of the pump laser is adjusted by a polarizing beam splitter (PBS) and a half-wave plate (HWP1). Orthogonally polarized photon pairs are generated by a ppKTP crystal in a Sagnac interferometer composed of a PBS and half-wave plates (HWP2, HWP3).  HWP2 works only for down-converted photons (810nm) and rotates the polarization to $90^{\circ}$. The reason for introducing HWP2 is to make the 1st and 2nd down-converted photons indistinguishable by compensating for the walk-off effect on the crystal. HWP2 is also used to modulate the relative phase between the pump light and the down-converted light by tilting the angle of HWP2. HWP3 works for both pump light (405 nm) and down-converted photons (810 nm) and rotates the polarization to $90^{\circ}$ and convert the polarization state of the pump light to horizontal to generate down-converted photons. 

A 10 mm ppKTP crystal (10$\mu$m periodically poled KTP crystal, AR coating for 405/810nm, RAICOL CRYSTAL ltd.) is set with the temperature about $28.5^\circ C$ for the down-conversion under degenerate condition of photon pairs with a center wavelength of 810nm. A 5mm ppKTP crystal is used to compensate for the walk-off effect caused by the birefringence property. A dichroic mirror (DM) separates the pump laser and down-converted photons. The output photons are reflected by the PBS and a DM and counted by detectors after passing through quarter-wave plates (QWPs), polarizers (POLs), and band-pass filters (BPFs) with a center wavelength of $810 \pm 3$ nm. The directors are standard single-photon counting modules (SPCMs) based on a Geiger-mode single-photon avalanche diode (Count-100C-FC, LASER COMPONENTS GmbH). We directly measure the pump power dependence of the single and coincidence counts of the down-converted photons from the 10 mm ppKTP crystal to estimate the actual photon-pair production rate with our source. The photon-pair production rate is defined as $N_{\rm pair}={N_S}^2 / N_C $ where  $N_S$ is the single count rate per unit input power and  $N_C$ is the coincidence count rate per unit input power \cite{tanzilli01}. 

This rate is a measure to estimate how much the produced photons contribute to the coincidence detections in the measurement. Note, the quantum efficiencies of detectors do not affect for $N_{\rm pair}$ from the definition. In our experiment, $N_S =(8.3 \pm 0.3)\times10^5$ Hz/mW, and  $N_C = 11.3 \pm 0.4$ Hz/mW. Therefore the rate of direct photon-pair production from the crystal is estimated as $N_{\rm pair} =(6.1 \pm 0.5)\times10^6 $ Hz/mW. If we simply excite the same crystal with twice the power,  $N_{\rm pair}$ is given by $2 \times {N_S}^2 / N_C $  because $N_S$ and $N_C$ increase linearly with the pump power. If we prepare the setup to stimulate the photon pairs with optimal phase relation as shown in Figure \ref{fig2},  $N_{\rm pair}$ is given by almost constant value because the second-order count rate becomes as $\sim 4 \times N_C$ in principle. These numbers are comparable to the photon-pair production rate given by the quantum interferometric scheme of $N_{\rm pair} \sim1.3\times10^7$ Hz/mW \cite{terashima18}, and also to the photon-pair production rate for a similar high-brightness degenerate entangled source of $N_{\rm pair} \sim3.9\times10^7 $ Hz/mW \cite{jabir}.  We could do similar discussion about higher-order photon production rate by defining $n$-th photon production rate as $N_{\rm n-photon}={N_S}^n / N_C $. 

The setup corresponds to the $N=2$ case of the schematic shown in Fig.  \ref{fig1}. The pump light of 0.9 mW power and 405 nm wavelength from a laser diode can be set with an arbitrary linear polarization state by PBS and HWP1. Orthogonally polarized photon pairs are generated in collinear spatial modes by exciting a ppKTP crystal twice. According to Eq. (\ref{pn}), the probability $P_2$ of generating polarization-entangled photon pairs depends on the phase $\theta$. In contrast, the probability $P_1$ of generating polarization-entangled photon pairs by exciting the same crystal once does not depend on the phase. The probability ratio is given by $P_2/P_1=2(1+\cos \theta)$. Therefore the maximum $P_2$ is expected to be four times larger than $P_1$ when the phase $\theta$ is set to zero.

\begin{figure}[h]%
\centering
\includegraphics[width=0.5 \textwidth]{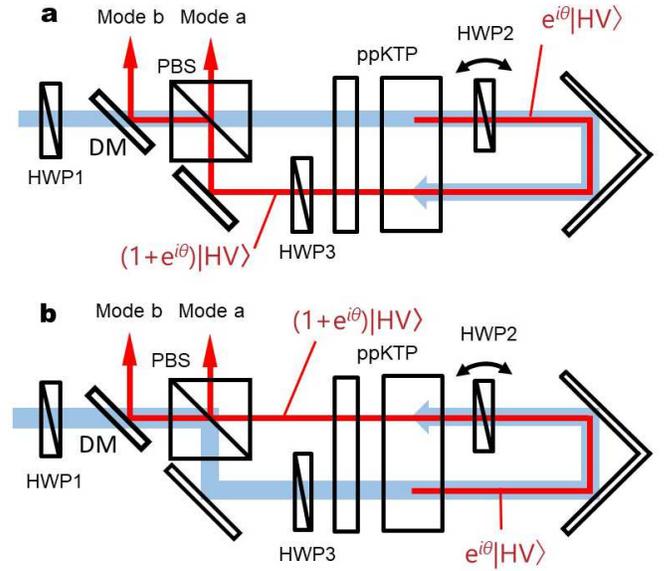}
\caption{Schematic of integrated double-pass polarization Sagnac interferometer using a polarizing beam splitter (PBS) and half-wave plates (HWP2, HWP3)  for (a) clockwise and (b) counterclockwise directions. Blue and red characters show the horizontal (H) or vertical (V) polarization state of the pump laser and down-converted photons, respectively.}\label{fig3}
\end{figure}

\begin{figure}[h]%
\centering
\includegraphics[width=0.5 \textwidth]{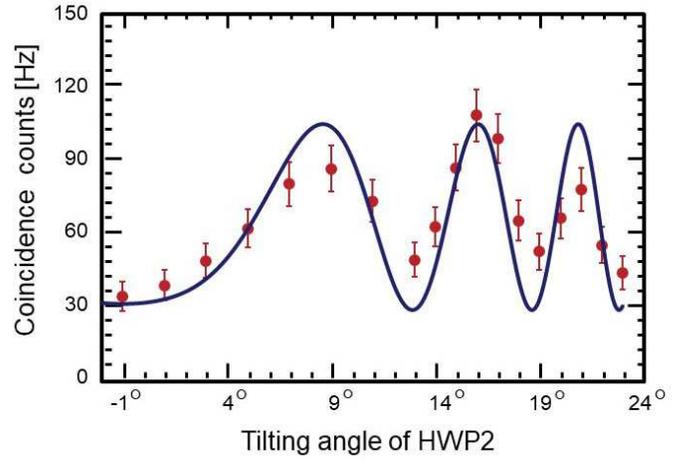}
\caption{Plot of coincidence counts of output photons produced by the clockwise process for different tilting angles of HWP2.}\label{fig4}
\end{figure}

First, the pump laser is set to 405nm wavelength with a horizontally polarized state, and the pump light propagates in the clockwise direction as shown in Fig. \ref{fig3}a. Under this condition, the horizontally polarized down-converted photons are output to mode A, and the vertically polarized down-converted photons are output to mode B.  By changing the tilting angle of HWP2, coincidence measurements for the output photons are performed using SPCMs and a computer (PC). We used QWPs, POLs, and BPFs for the polarization and band-pass filters to minimize the number of linear optics components in the setup. Results of the coincidence measurement are shown as Fig. \ref{fig4}. The count rate of the figure is shown with the scale of actual coincidences. The actual coincidences in the measurement are even smaller than $N_C = 11.3 \pm 0.4$ Hz/mW because we used pinholes to overlap the spatial optical modes of generated photons.

When we remove HWP2 from the setup, the down-converted photons are not stimulated because the birefringent effect of the ppKTP crystal makes the down-converted photon pairs on  the same optical modes distinguishable. Therefore, the experimental value of $P_1$ can be estimated as half the value of the probability using the setup without HWP2. From the comparison with the experimentally estimated  $P_1$ value, the maximum probability ratio of $P_2/P_1$ is estimated as $3.4 \pm 0.1$. \\

We also estimated the maximum probability ratio by simply fitting the expected function $2A\{1+B \cos (\Delta \phi+C) \}$. Here, $A, B,$ and $C$ are constant parameters to be estimated by fitting. $\Delta \phi$ is the variable for the function. The relation between the variable and the tilting  angle of HWP2 is derived in Appendix B. $P_1$ is given by the fitting parameter $A$, and the maximum $P_2$ is given by $2A(1+B)$. Therefore, the maximum probability ratio of $P_2/P_1$ is given by the parameter $B$ and estimated using $2(1+B)$ as $2.7\pm0.1$. \\

It is presumed that the difference in the numerical values of  $P_2/P_1$ estimated by the above two methods is due to the optical mode shift caused by the removal of HWP2 and also the reflection caused by the existence of HWP2. Either way, $P_2$ is more than double $P_1$ owing to the stimulation effect. \\

\begin{figure}[h]%
\centering
\includegraphics[width=0.5 \textwidth]{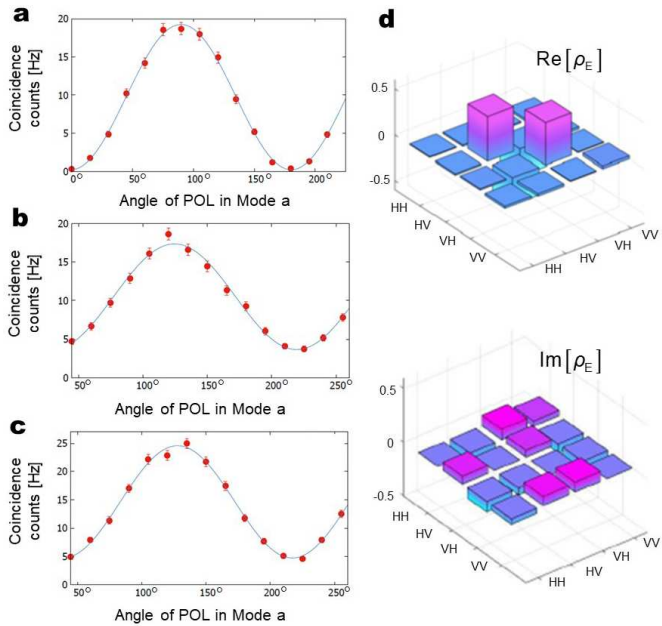}
\caption{Plot of coincidence counts under linear polarization basis when mode B POL angle is fixed at  (a) 0$^{\rm o}$ and  (b) 45$^{\rm o}$ and (c) under circular polarization basis. Solid lines are the best sinusoidal fits to the data. (d) Reconstructed two-photon density matrix with real and imaginary parts obtained by quantum state tomography.}\label{fig5}
\end{figure}

Under the condition giving the maximum probability ratio of $P_2/P_1$ with the setup shown in Fig. \ref{fig3}a, we set HWP1 at $-45^{\circ}$ to rotate the polarization state of the pump light to generate polarization-entangled photon pairs in the Bell state $ | {\Phi}_1^- \rangle = ( | H_a V_b \rangle -   | V_a H_b \rangle )/ \ \sqrt{2} $. Fig. \ref{fig5}a and b show the coincidence counts for generated photon pairs set in the ${\Phi}_1^- $ state under a linear polarization basis  as a function of the mode A POL angle when the mode B POL angle is fixed at 0$^{\rm o}$ and 45$^{\rm o}$, respectively. The count rates of the figures are shown with the scale of actual coincidences. We remove the QWPs in modes A and B for the measurement. The solid curves are the best sinusoidal fits to the corresponding data. The visibility of the fringe is defined by $(C_{\rm max}-C_{\rm min})/(C_{\rm max}+C_{\rm min}) $, where $C_{\rm max}$ is the maximum coincidence count and $C_{\rm min}$ is the minimum coincidence count. The visibilities estimated from the sinusoidal fits in Fig.  \ref{fig5}a and b are $0. 94\pm0.02$ and $0. 70\pm0.03$, respectively. 

Fig. \ref{fig5}c shows the coincidence count for the same state under a circular polarization basis as a function of the mode A POL angle when the mode B POL angle is fixed at 45$^{\rm o}$ and the angles of both QWPs are set at 0$^{\rm o}$. The estimated visibility is $0. 74\pm0.04$.  These reasonable visibilities are clear evidence of the nonclassical quantum interference given by the polarization-entangled state. These visibilities are currently limited, mainly because we use multimode fibers to collect the generated entangled photons since the tilting of HWP1 affects the overlapping of spatial modes.  The pump laser beam passed a nonlinear crystal two times for both the CW and CCW direction of the Sagnac interferometer. The visibility of interference fringes deteriorates because the height difference causes spatial mode mismatch on the output of the Sagnac interferometer. The visibility of interference fringes is expected to be higher when using a singlemode fiber due to the spatial mode-overlapping area of the CW and CCW beams. We verified the above effect by performing an additional experiment using a diode laser with the Sagnac interferometer \cite{terashima18,terashima19}.

We also measure the photon pairs by reconstructing the complete density matrix by quantum state tomography, which requires coincidence measurements for 16 combinations of polarization bases\cite{altepeter05}. The experimentally reconstructed real and imaginary parts of the two-photon polarization density matrix $\rho_\mathrm{E}$ are shown in Fig. \ref{fig5}d. The fidelity of the experimentally reconstructed density matrix to an ideal entangled state ${\Phi}_1^- $ is given by $F=\langle {\Phi}_1^- | \rho_\mathrm{E}| {\Phi}_1^- \rangle$. We estimate the fidelity as $0.88 \pm 0.01$. The result shows that the generated photon pairs are sufficiently in the $\phi^{-} $ state even after the stimulation by the scheme in a double-pass configuration.

\section{Discussion}
To further increase the stimulation emission probability of polarization-entangled photon pairs by more than four times, we need to increase the number of times to excite the nonlinear crystals more than three times. It is necessary to determine how to compensate for the walk-off effect caused by the birefringence of a nonlinear crystal for the expansion. One solution is to use down-converted photons generated by the type-0 phase matching process. As shown in the previous experiment, it is possible to generate unconditional and degenerated polarization-entangled photons by the phase matching process with high emission efficiency \cite{terashima18,terashima19}.   

The size of a commercially available nonlinear crystal is typically about a few centimeters square. It is possible to excite such crystals about ten times with the setup shown in Fig. \ref{fig1}b.  Therefore, we expect to increase the emission probability of unconditional polarization-entangled photons by more than two orders of magnitude compared with the conventional case using only typically available passive optical components. Under the current double-pass configuration for maximally four-times enhancement of polarization-entangled photon pairs, we can neglect the unwanted effects from higher-order multi photon events (Refer to Appendix A). Under much higher multiple enhancement, higher-order multi photon events might increases the error rates in quantum applications \cite{brassard00}. There are several proposals to moderate such unwanted effects and to improve the signal to noise ratio using passive temporal multiplexing schemes \cite{ma11, broome11}. Although the present visibility and entanglement fidelity is not enough for the practical level of these quantum applications, the fabrication of waveguide structures on nonlinear crystals can improve these values in future experiments \cite{levine11}.

\section{Summary}

We proposed a scale-up scheme to stimulate polarization-entangled photon pairs using a resonator with only passive optical components. We theoretically showed how it is enhancing the stimulated emission of polarization-entangled photons and also performed a proof-of-principle experimental demonstration of our scheme in a double-pass configuration. Experimental results showed that the probability of generating the polarization-entangled photon pairs is higher than that in the conventional case, which match the theoretical expectation with reasonable visibilities and fidelity even after the stimulation process. Our theoretical proposals and experimental results will lead to the realization of highly efficient and bright quantum entangled photon sources required in quantum information technologies. This work was supported by JST Grant-in-Aid for Scientific Research (C) Grant Number 21K04931.

\appendix

\section{General description of stimulated emission}

In this section, we show that our scheme enhances the probability of maximally entangled state generation by a factor of $O(N^{2M})$ 
with the number of entangled photons $2M (M=1,2,3 \cdots)$ and the number of times $N$ that a single laser pulse passes through the ppKTP crystal. 
Here, the maximally entangled state of $2M$ photons is defined in Ref.~\cite{lamas01} as 
\begin{align} 
	\ket{{\Phi}^-_M} =\dfrac{1}{\sqrt{M+1}}\sum_{k=0}^{M}(-1)^{k}\ket{M-k, k; k, M-k}; \ (M \in \mathbb{N}), \label{eq:ment}
\end{align} 
where $n_{cP}$ ($c = a, b; \ P = H, V$) of the state $\ket{n_{aH}, n_{aV}; n_{bH}, n_{bV}}$ in Eq.~(\ref{eq:ment}) represents the number of $P$-polarized photons in the mode $x$. 

Let us begin by considering the time evolution of the system. 
As described in Sec. II, the total Hamiltonian of the system can be expressed as 
\begin{align}
\hat{H}_{\mathrm{out}} 
	= {\hat{H}}_{\mathrm{CW}} - {\hat{H}}_{\mathrm{CCW}} 
	={\kappa} \sum_{m=0}^{N-1} e^{im \varphi}\left(
	\hat{a}^{\dagger}_{H} \hat{b}^{\dagger}_{V} - \hat{a}^{\dagger}_{V} \hat{b}^{\dagger}_{H} 
	\right) \nonumber \\
    + \mathrm{h.c.} \nonumber \\
 \label{eq:Hout1}
\end{align}
Here, we note that the operators ${\hat{H}}_{\mathrm{CW}}$ and ${\hat{H}}_{\mathrm{CCW}}$ are defined by Eqs.~(1) and (2) of 
Sec.~II, respectively. 
The Hamiltonian in Eq.~(\ref{eq:Hout1}) can be simplified to 
\begin{align}
	\hat{H}_{\mathrm{out}}=\kappa A \left(
	\hat{a}^{\dagger}_{H} \hat{b}^{\dagger}_{V} - \hat{a}^{\dagger}_{V} \hat{b}^{\dagger}_{H} 
	\right) 
	+ \mathrm{h.c.} , \label{eq:Hout2}
\end{align} 
where the constant $A$ is defined by 
\begin{align}
	A \equiv \sum_{m=0}^{N-1}e^{im \varphi} 
	=1+e^{i \varphi }+ \cdots + e^{i(N-1)\varphi }
	=\dfrac{1-e^{iN \varphi }}{1-e^{i\varphi }}. \label{eq:defA}
\end{align} 
The time-evolution operator of ${\hat{H}}_{\mathrm{out}}$ in Eq.~(\ref{eq:Hout2}) is given by 
\begin{align}
	\hat{U} (t) = \exp \left [ -\dfrac{i\hat{H}_{\mathrm{out}}t}{\hbar} \right ]
	= \exp \left [ - i A \tau {\hat{L}}_{+} - i A^* \tau {\hat{L}}_{-} \right ], \label{ut}
\end{align}
where we define $\tau \equiv \kappa t/\hbar$ and 
${\hat{L}}_{+} \equiv \hat{a}^{\dagger}_{H} \hat{b}^{\dagger}_{V} - \hat{a}^{\dagger}_{V} \hat{b}^{\dagger}_{H}= {\hat{L}}_{-}^{\dagger}$. 
Then, the output state can be expressed as Eq.~(3) in Sec. II, or equivalently, $\ket{{\Psi}_{\mathrm{out}}} = \hat{U} (t) \ket{0}$ 
with the vacuum state $\ket{0}$. 

The probability $P_{M,N}$ of obtaining $| {\Phi}^-_M \rangle$ defined in Eq.~(\ref{eq:ment}) is given by 
$P_{M,N} = | \langle {\Phi}_M^- | \hat{U} (t) | 0 \rangle |^2$.
To see how $P_{M,N}$ depends on $M$ and $N$, we simplify the time-evolution operator in Eq. (A5). $\to$ on $M$ and $N$, we apply the disentangling theorem \cite{Kok,levine11} to the time-evolution operator in Eq. (A5), since the operators ${\hat{L}}_+$ and ${\hat{L}}_-$, together with ${\hat{L}}_0 \equiv \frac{1}{2} [{\hat{L}}_-, {\hat{L}}_+]$, satisfy the commutation relations of the special unitary algebra $\mathrm{su} (1,1)$, that is, $[{\hat{L}}_0, {\hat{L}}_{\pm}] = \pm {\hat{L}}_{\pm}$ and $[{\hat{L}}_+, {\hat{L}}_-] = -2{\hat{L}}_0$. Then, we obtain
\begin{align}
   \hat{U} (t) = \exp {[-i (\tanh |A\tau|) {\hat{L}}_+ ]}  
   \exp {[-2\ln (\cosh |A\tau|) {\hat{L}}_{0}]} \nonumber \\
   \times \exp {[-i (\tanh |A\tau| ) {\hat{L}}_{-}]}, 
   \label{ut2}
\end{align} 
which acts on $\ket{0}$ to generate $\ket{{\Psi}_{\mathrm{out}}}$ as 
\begin{align}
  \ket{{\Psi}_{\mathrm{out}}} 
  &= \dfrac{1}{\cosh^{2}|A\tau|} \exp [ -i (\tanh |A\tau| ) {\hat{L}}_{+} ] \ket{0} \nonumber \\ 
  &= \dfrac{1}{\cosh^{2}|A\tau|}
  \sum_{n=0}^{\infty} (-i)^{n}\tanh^{n} |A\tau| \nonumber \\
  & \times \sum_{l=0}^{n}(-1)^{l}\ket{n-l, l; l, n-l}, \label{eq:sout}
\end{align} 
since $\exp [-i (\tanh |A\tau|) {\hat{L}}_{-}] \ket{0} = \ket{0}$ and 
\begin{align} 
	{\hat{L}}_0 = \frac{1}{2} \left ( {\hat{a}}_H^{\dagger} {\hat{a}}_H + {\hat{a}}_V^{\dagger} {\hat{a}}_V 
	+ {\hat{b}}_H^{\dagger} {\hat{b}}_H + {\hat{b}}_V^{\dagger} {\hat{b}}_V \right ) + 1, 
\end{align} 
implying that $\exp [-2\ln (\cosh |A\tau|) {\hat{L}}_{0}] \ket{0} = (\cosh |A \tau |)^{-2} \ket{0}$. 
From Eqs.~(\ref{eq:ment}) and (\ref{eq:sout}), we obtain 
\begin{align} 
	P_{M,N} = | \langle {\phi}_M^- | \hat{U} (t) | 0 \rangle |^2 = \frac{M+1}{{\cosh}^4 |A\tau |} {\tanh}^{2M} |A\tau |. 
	\label{eq:PMN}
\end{align} 
Note that Eq.~(\ref{eq:PMN}) implicitly depends on $N$ through $A$ defined in Eq.~(\ref{eq:defA}). 

Now, let us maximize $P_{M,N}$ in Eq.~(\ref{eq:PMN}) with respect to $\varphi$. 
We define $u \equiv {\tanh}^2 |A \tau|$ and express $P_{M,N}$ as a function of $u$ as 
\begin{align} 
	P_{M,N} = (M+1) (1 - u)^2 u^M \equiv f(u). \label{eq:deff} 
\end{align} 
The first derivative of Eq.~(\ref{eq:deff}) has three zeros at $u = 0$, $\frac{M}{M+2}$, and $1$, which means that 
$f(u)$ monotonically increases in the region $u \in [0, \frac{M}{M+2}]$ and takes its maximum value at 
$u = \frac{M}{M+2}$. 
On the other hand, $u$ should satisfy $u \approx |A \tau |^2 \ll \frac{M}{M+2}$, since $\tau$, which is given by 
the square root of the conversion efficiency of the nonlinear optical crystal, is very small as $\tau \ll O(1)$ and 
\begin{align} 
	|A| = \left | \frac{\sin {\frac{N \varphi}{2}}}{\sin {\frac{\varphi }{2}}} \right | \leq N \sim O(1), \label{eq:|A|}
\end{align} 
where the equality holds in the limit $\varphi \to 0$. 
Therefore, $P_{M,N}$ can be approximated by 
\begin{align} 
	P_{M,N} = (M+1) {\tau}^{2M} {\left | \frac{\sin {\frac{N \varphi}{2}}}{\sin {\frac{\varphi }{2}}} \right |}^{2M}, 
\end{align}
which is maximized at $\varphi = 0$ to be 
\begin{align} 
	P_{M,N} = (M + 1) | N \tau |^{2M}, \label{eq:Pmax}
\end{align} 
up to the order $O(|N \tau |^{2M})$.  From Eq.~(\ref{eq:PMN}), $P_{2,N}/P_{1,N}$, the ratio of the probability of detecting four entangled photons to the probability of detecting two entangled photons, depends on $\tanh^{2} |A\tau|$. Under the current condition of $\tau \ll O(1)$ and $|A|=N=2$, $\tanh^{2} |A\tau|$ is very small due to $|A\tau| \ll 1$. Therefore, it is not necessary to consider the effect caused by more than two photons.

\section{Manipulation of relative phase using a half-wave plate}

When light with wavelength $\lambda$ passes through a transparent dielectric with length $L$ and index $n$, the phase of the light changes by $2\pi n/\lambda$. When a dielectric plate is placed with a tilting angle $\alpha$ from the normal vector as shown Fig. \ref{fig6}, the relation between the injection angle and the refraction angle is given by Snell's law as $\sin \alpha / \sin \beta =n$. Therefore, when the light passes through a transparent dielectric under the configuration as shown Fig.  \ref{fig6}, the phase of the light changes as follows:

\begin{figure}[h]%
\centering
\includegraphics[width=0.5 \textwidth]{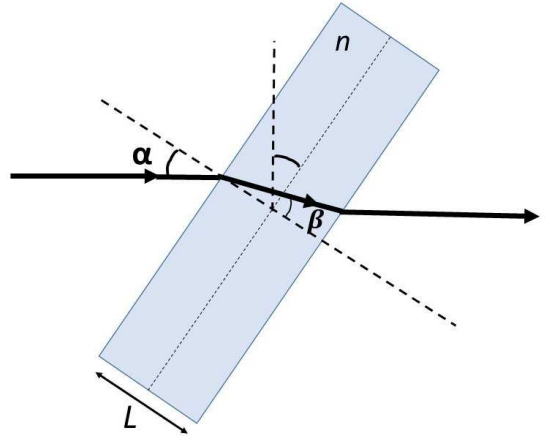}
\caption{Schematic illustration of our relative phase modulation by a half-wave plate (HWP) constructed with a thin birefringent material layer sandwiched by N-BK7 glass plates with a tilting angle.}\label{fig6}
\end{figure}

\begin{equation}
\label{phi_1}
\phi=\frac{2\pi n}{\lambda} \frac{L}{\cos \beta} \equiv \frac{2\pi n^2}{\lambda} \frac{L}{\sqrt{n^2-\sin^2 \alpha}}.
\end{equation}

When a half-wave plate (HWP2) is placed with a tilting angle $\alpha$, the phase of the input light is given by the same relation. When the pump light with wavelength $\lambda_p$ and down-converted photon pairs with wavelengths $\lambda_s$ and $\lambda_i$ are simultaneously injected into HWP2, the phases of these wavelengths $\phi_s, \phi_i$, and $\phi_p$ change as follows:

\begin{equation}
\label{phi_2}
\phi_X= \frac{2\pi n_X^2}{\lambda_X} \frac{L}{\sqrt{n_X^2-\sin^2 \alpha}} \ (X=p, s, i).
\end{equation}

Here $n_s, n_i$, and $n_p$ are the refractive indexes for the light with wavelengths $\lambda_s, \lambda_i$, and $\lambda_p$ respectively. The relative phase change caused by the dispersion between the inputs is given as $\Delta \phi = \phi_p -\phi_s -\phi_i$. The relation $1/\lambda_p=1/\lambda_s+1/\lambda_i$ is given by the energy conservation law. When the down-converted photons are generated under degenerated condition, the relation between the wavelengths are given as $2\lambda_p=\lambda_s=\lambda_i$. Therefore, the relation between $\Delta \phi$ and $\alpha$ becomes

\begin{equation}
\label{phi_3}
\Delta \phi= \frac{2\pi L}{\lambda_p} \left\{ \frac{n_p^2}{\sqrt{n_p^2-\sin^2 \alpha}} - \frac{n_s^2}{\sqrt{n_s^2-\sin^2 \alpha}}  \right\}.
\end{equation}

In our experiment, the down-converted photons are generated under the degenerated condition with the wavelength of the pump light $\lambda_p=405$ nm. HWP2 is composed of a few micron-meter order thin birefringent material layer.  N-BK7 glass plates of a few-millimeter-order thickness sandwich the birefringent material layer. Therefore, the refractive indexes of N-BK7 glass practically determine the dispersion effects for the input lights. We used refractive indexes of $n_p \sim 1.53$, $n_s \sim 1.51$, and $L \sim 3$ mm for the fitting of data shown in Fig.  \ref{fig4}.


\end{document}